\documentclass[aps,prb,twocolumn,a4]{revtex4-1} 
\pdfoutput=1
\usepackage{amsmath}
\usepackage[dvipsnames]{xcolor}
\usepackage{graphicx}
\usepackage{setspace}
\usepackage{comment}
\usepackage{hyperref}
\usepackage{float}
\usepackage[normalem]{ulem} 
\usepackage{upgreek}

\newcommand{\LTO}{LaTiO$_3$}
\newcommand{\LTOO}{LaTiO$_3$ }
\newcommand{\LTOov}{LaTiO$_3$(O$_{\rm V}$)}
\newcommand{\LTOovv}{LaTiO$_3$(O$_{\rm V}$) }
\newcommand{\tg}{$t_{2g}$}
\newcommand{\tgg}{$t_{2g}$ }

\newcommand{\egg}{$e_g$ }
\newcommand{\ov}{O$_{\rm V}$}
\newcommand{\ovv}{O$_{\rm V}$ }

\begin{document}

\title{DFT+DMFT study of oxygen vacancies in a Mott Insulator}

\author{Jaime Souto-Casares}
\affiliation{Materials Theory, ETH Z\"urich, Wolfgang-Pauli-Strasse
  27, 8093 Z\"urich, Switzerland}
\author{Nicola A. Spaldin}
\affiliation{Materials Theory, ETH Z\"urich, Wolfgang-Pauli-Strasse
  27, 8093 Z\"urich, Switzerland}
\author{Claude Ederer}
\affiliation{Materials Theory, ETH Z\"urich, Wolfgang-Pauli-Strasse
  27, 8093 Z\"urich, Switzerland}

\date{\today}
\begin{abstract}

Oxygen vacancies are a common source of excess electrons in complex oxides. In Mott insulators these additional electrons can induce a metal-insulator transition (MIT), fundamentally altering the electronic properties of the system. Here we study the effect of oxygen vacancies in \LTO, a prototypical Mott insulator close to the MIT. We show that the introduction of oxygen vacancies creates a vacancy-related band immediately below the partially filled Ti-\tgg bands. We study the effect of this additional band on the Mott MIT using a combination of density functional theory and dynamical mean-field theory (DFT+DMFT), employing a minimal correlated subspace consisting of effective Ti-\tgg orbitals plus an additional Wannier function centered on the vacancy site. We find that the Mott insulating state in \LTOO is robust to the presence of the vacancy band, which remains fully occupied even in the presence of a local Coulomb repulsion, and therefore does not cause a doping of the Mott insulator.

\end{abstract}

\maketitle


Point defects are an unavoidable feature in perovskite oxides at finite temperature. Among them, oxygen vacancies~\cite{nota_OVnotation} (O$_{\rm V}$) are believed to play a key role in a variety of emergent phenomena, such as superconductivity~\cite{Cava1987}, the establishment of an interfacial two-dimensional electron gas~\cite{PhysRevB.75.121404}, magnetoresistance~\cite{Kormondy2018}, or blue-light emission at room temperature~\cite{Hwang2005,Crespillo2017}. Despite their relevance, a complete picture of the effect of oxygen vacancies in complex functional oxides is still lacking, in part due to the strong coupling of multiple degrees of freedom (structural as well as electronic) in these systems~\cite{ganduglia2007oxygen,Eckstein2007,PhysRevB.94.241110}.
While it is well known for some perovskite oxides that O$_{\rm V}$'s create defect states inside the energy gap, the itinerant or localized nature of these states remains the key open question in this field~\cite{PhysRevB.90.085202,PhysRevLett.111.217601,PhysRevB.93.121103}.

The Mott metal-to-insulator transition (MIT) is an intriguing phenomenon in complex oxides, where electronic correlation effects play a central role (for a review, see Ref.~\onlinecite{RevModPhys.70.1039}). Although many aspects of the Mott MIT, for realistic materials, are not fully understood, the use of such materials in novel functional applications within the emergent field of {\it Mottronics}~\cite{Mannhart2012,Inoue2008} has immediate relevance. Oxygen vacancies can have a potentially large effect on the MIT, either by changing the stochiometry, disordering the lattice, or introducing chemical strain through lattice expansion~\cite{PhysRevB.88.054111}.

Within a Mott insulator, provided that the vacancy-induced band is partially occupied, one can expect that correlation effects would penalize double occupancy and instead change the valence state of neighboring cations. This could then destabilize the Mott insulating state, similarly to a doped Mott insulator within the Hubbard model, where the doping breaks the commensurability between  the number of electrons and the number of sites~\cite{0295-5075-118-1-17004,PhysRevB.79.115119}. Thus, in contrast to an uncorrelated semiconductor, where the amount of carriers is essentially proportional to the amount of defects, in a Mott insulator the defect-induced doping could fundamentally alter the electronic state (by causing a MIT), thereby effectively transforming \emph{all} valence electrons into carriers. This potential MIT caused by the oxygen deficiency has not been widely explored, due to the complicated interpretation of the experimental results\cite{PhysRevB.93.121103} and also to the lack of a suitable theoretical framework.

In this work, we directly address the latter point by using DFT+DMFT~\cite{georges1996dynamical,anisimov1997first,held2007electronic,PhysRevLett.75.105,Pavarini,PhysRevLett.87.067205,pavariniPRL} to investigate the Mott insulating state of the prototypical Mott-insulator \LTOO in the presence of oxygen vacancies. \LTOO has an orthorhombically-distorted $Pnma$ perovskite structure ($a^-b^+a^-$ distortion in Glazer's notation\cite{Glazer:a09401}) and a reported optical gap of $\sim0.2$ eV\cite{PhysRevB.51.9581}. The Ti atoms are in a Ti$^{+3}$ state, with one electron in the \tgg orbitals ($d^1$). The relative simplicity of its electronic structure and its proximity to the MIT from the insulating side makes \LTOO a perfect model system to study the stability of the Mott insulating state.


To obtain accurate geometries and bandstructures, we perform standard DFT calculations using the projector-augmented wave (PAW) method, as implemented in the ``Vienna ab-initio simulation package'' (VASP)~\cite{Kresse/Furthmueller_CMS:1996,Kresse/Joubert:1999}, version 5.4.1, together with the GGA-PBE exchange-correlation functional~\cite{PBE}. The valence configurations of the PAW potentials used are La($5s6s5p5d$), Ti($3p4s3d$) and O($2s2p$). The La PAW potential includes the empty $4f$ states, which form a set of narrow bands between the Ti-\tgg and the \egg bands. We apply $U=6$ eV on the DFT$+U$ level~\cite{PhysRevB.57.1505} to shift these La-$f$ orbitals up in energy reducing the entanglement with the Ti-\tgg bands. To accommodate the $Pnma$ structure of \LTO, we use a $20$-atom unit cell, $19$-atom for the \ov-defective systems LaTiO$_{2.75}$ (vacancy concentration of 8.3$\%$). Converged results are obtained by sampling the Brillouin zone with a $11\times 9\times 11$ $\Gamma$-centered k-mesh and using a plane wave energy cutoff of $900$ eV. All structural degrees of freedom are relaxed until all forces are smaller than $10^{-4}$ eV/\AA. All calculations are performed without considering spin polarization.

The low-energy correlated subspace for the DMFT calculations is then constructed using a basis of maximally localized Wannier functions (MLWF)~\cite{lechermann2006dynamical}, employing the Wannier90 code~\cite{mostofi2014updated}. We use the TRIQS/DFTTools package~\cite{PARCOLLET2015398,TRIQS/DFTTools} to perform the paramagnetic DMFT calculations. An effective impurity problem is solved for each inequivalent Ti site plus the vacancy site using the TRIQS/CTHYB solver~\cite{Seth2016274}, while the different impurity problems are coupled through the DMFT self-consistency. The local interaction is modeled using the Hubbard-Kanamori parametrization with spin-flip and pair-hopping terms included, and Held's formula~\cite{nota_dc} is used to compute the double-counting term. All calculations are performed at room temperature, $\beta=(k_BT)^{-1}=40$ eV$^{-1}$, with a Hund's coupling of $J=0.64$ eV for the Ti sites, and without charge self-consistency. The values for the Hubbard $U$ are varied to analyze the effect on the electronic properties. Real frequency spectral functions, $A(\omega)$, are obtained from the local Green's functions in imaginary time, $G(\tau)$, using the Maximum Entropy algorithm\cite{Bryan1990}. The spectral weight around the Fermi energy, $\bar{A}(0)$, is calculated from the impurity Green's function as $\bar{A}(0)=-\beta/\pi\, G(\beta/2)$. 


\begin{figure}
	\includegraphics[width=0.45\textwidth]{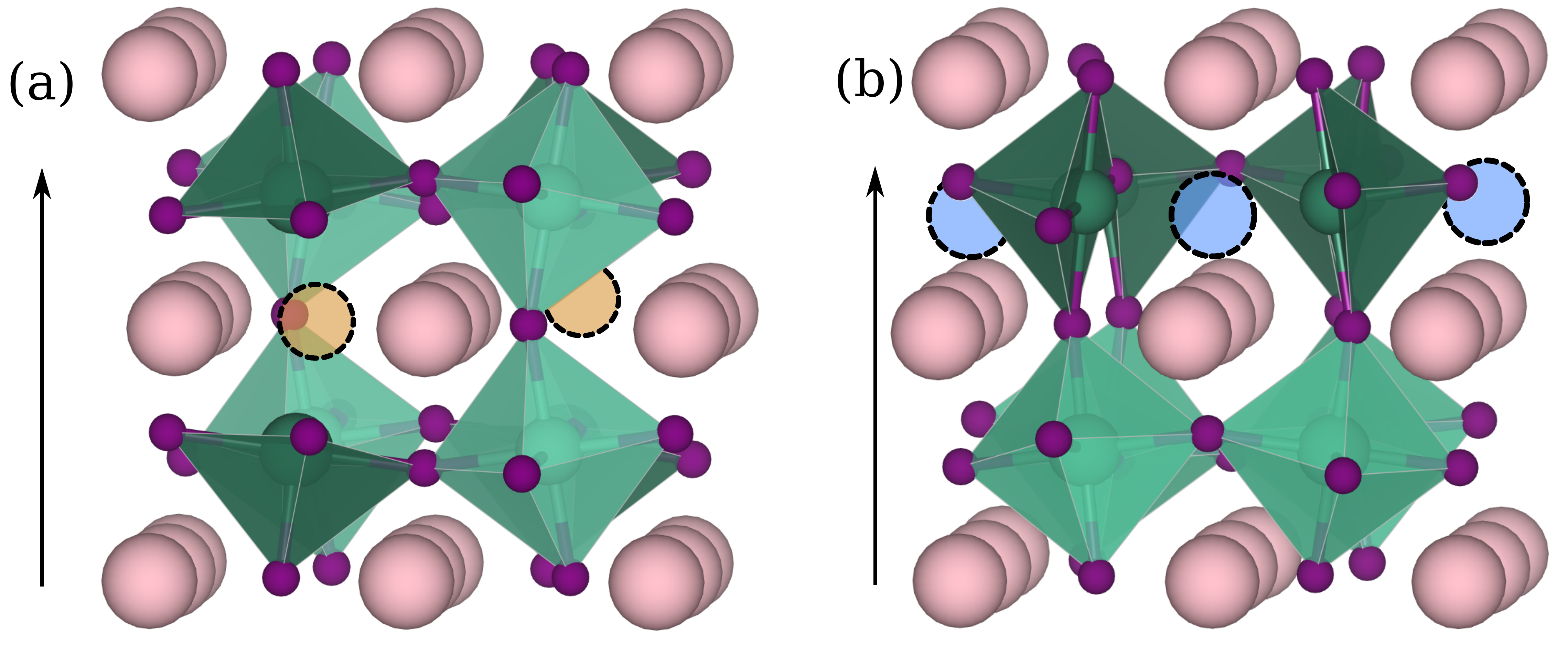}
	\caption{(Color online) Calculated LaTiO$_{2.75}$ structures with oxygen vacancies in the two inequivalent positions (see text): (a) axial, and (b) planar. Pink, green, and purple spheres represent La, Ti, and O atoms respectively. The darker (incomplete) octahedron surrounds the Ti {\it next} to the \ov, while the brighter one surrounds the Ti {\it farther} from it. Orange and blue spheres represent the position of the missing oxygens in the two cases. Arrows point along the direction of the long orthorhombic axis of the original $Pnma$ structure.}
	\label{fig:OV_geo}
\end{figure}

In the {\it Pnma}-distorted perovskite structure, there are two inequivalent sites for the oxygen atoms (Wyckoff positions 4c and 8d), hence two different vacancy sites. We denote as {\it axial/planar vacancy} the configurations where the missing oxygen belongs to an O-Ti bond parallel/perpendicular to the long orthorhombic axis (see Figs.~\ref{fig:OV_geo}a and \ref{fig:OV_geo}b, respectively). The oxygen vacancy lowers the symmetry of the system compared to pure $Pnma$ \LTO. In the axial case, the resulting space group (for the 20-atom cell used here) is $Pm$, with two inequivalent Ti sites, one next to the vacancy (its oxygen octahedron is missing one vertex) and one farther away (its oxygen octahedron is still intact). The planar case lowers the symmetry even further ($P1$), rendering all four Ti inequivalent. However, the difference between the two octahedra next to O$_{\rm V}$ (and between the two farther from O$_{\rm V}$) after the relaxation is very small and thus negligible. 

We start by relaxing the structure of pure \LTOO using DFT-GGA, for which the calculated geometry shows good agreement with available experiments and previous computer simulations~\cite{PhysRevB.89.161109,PhysRevB.68.060401,EITEL198695}. For the cases with \ov , the atomic positions are allowed to relax within the cell volume and shape of the pure \LTOO case. Test calculations showed that the effects of performing a full relaxation are small (the volume increases by $0.08\%$) and irrelevant for the current discussion. 

\begin{figure}
	\includegraphics[width=0.5\textwidth]{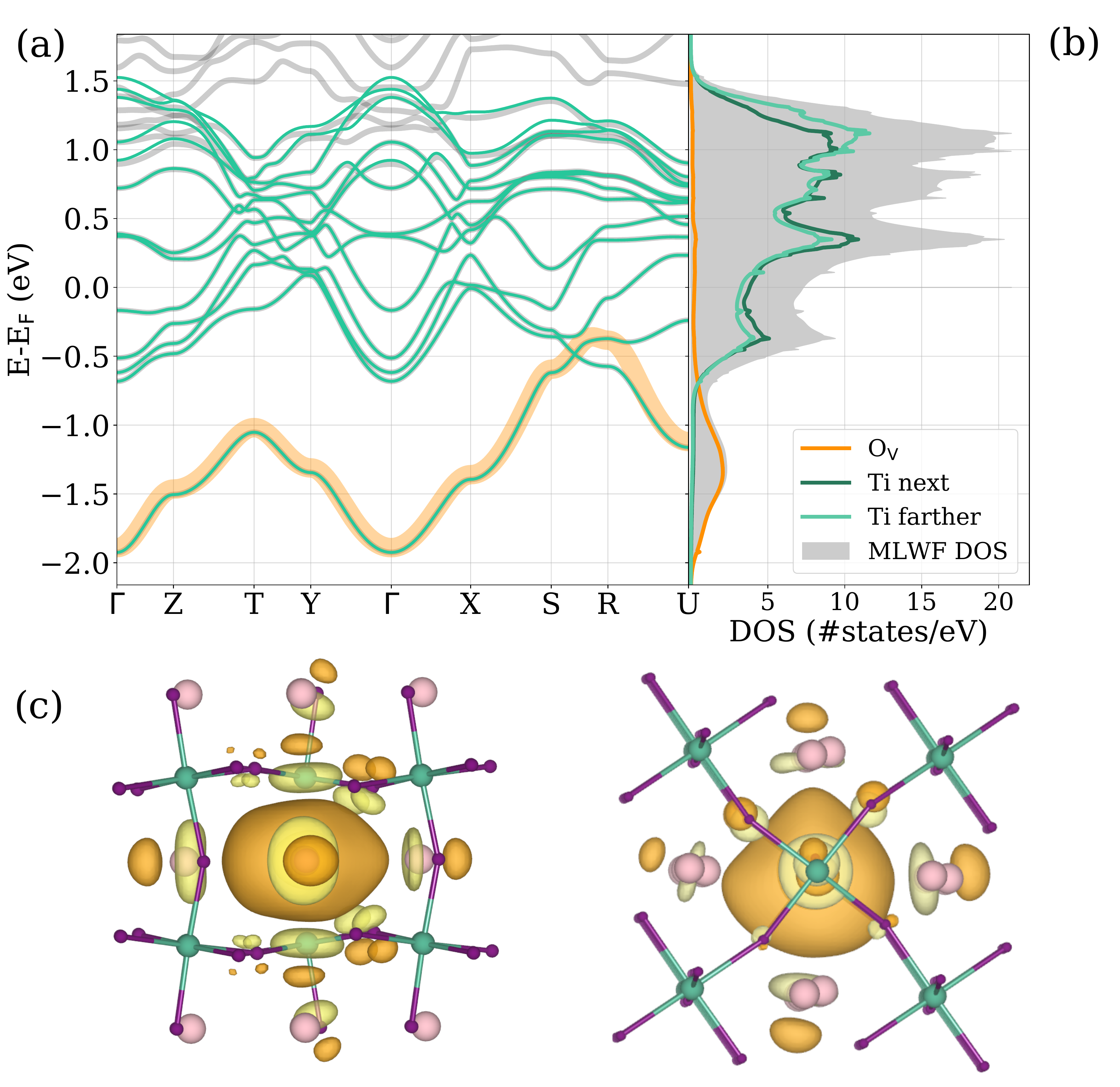}
	\caption{(Color online) (a) DFT bandstructures of LaTiO$_{2.75}$ for the axial vacancy configuration. DFT bands are shown in gray, while the MLWF bands are superimposed with green solid lines. The vacancy band, lying ${\sim}1.5$ eV below $E_{\rm F}$, is highlighted with a thicker line representing the weight of the O$_{\rm V}$ MLWF. (b) MLWF-projected density of states (DOS) for the Ti-\tgg and O$_{\rm V}$-MLWFs. (c) Real-space representation of the O$_{\rm V}$-MLWF for axial vacancy, with side (left) and top (right) views with respect to the long $Pnma$ axis of the stoichiometric system.}
	\label{fig:OV01_dft}
\end{figure}

The calculated DFT bandstructure of pure \LTOO shows a band with predominant Ti-\tgg character around the Fermi energy, and a Ti-\egg band slightly higher in energy, with some level of entanglement between them in the vicinity of the $\Gamma$ point (see, e.g., Ref.~\onlinecite{PhysRevB.89.161109,Pavarini}). The O-$p$ derived band lies ${\sim}4$ eV below the Fermi energy. Predicting metallic behavior, spin-unpolarized DFT-GGA fails to properly reproduce the correlated nature of the $d$ electrons of \LTO.

The bandstructure plot for the axial O$_{\rm V}$ configuration is shown in Fig.~\ref{fig:OV01_dft}a. The most prominent difference with respect to the structure of pure \LTOO is the presence of a band ${\sim}1.5$ eV below $E_{\rm F}$. 
A similar feature has been attributed to the O$_{\rm V}$ in other perovskite systems, e.g., in SrTiO$_{3-\delta}$~\cite{PhysRevB.90.085202,PhysRevLett.111.217601} or SrVO$_{3-\delta}$~\cite{PhysRevB.93.121103}. In the axial configuration, the vacancy band shows a strong dispersion, crossing the Ti-\tgg band around the $R=(\tfrac{1}{2},\tfrac{1}{2},\tfrac{1}{2})$ k-point, whereas for the planar configuration (not shown here), the vacancy band remains detached from the Ti-\tgg bands across the whole Brillouin zone. We note that the high vacancy concentration used in the present calculations (${\sim}8.3$\%) likely produces an overestimation of the \ov-band dispersions.

In the following, we investigate how the presence of this O$_{\rm V}$ band affects the Mott-insulating character of \LTOov. As shown in previous work~\cite{Pavarini,PhysRevB.89.161109,RevModPhys.70.1039}, a good description of the low-energy physics of an early transition metal oxide like \LTOO is obtained by including only the Ti-$t_{2g}$-dominated bands around the Fermi level into the correlated subspace used for the DMFT calculation. In the present case, we extend this correlated subspace to also include the \ovv band slightly below these $t_{2g}$ bands. Thus, we construct MLWFs for the 12 Ti-\tgg bands plus the \ovv band by defining an appropriate energy window and using initial projectors corresponding to 3 $t_{2g}$ orbitals located at each of the 4 Ti sites within the unit cell and an additional $s$ orbital trial projector centered on the vacancy site. Fig.~\ref{fig:OV01_dft}a shows the good agreement between the DFT bands and the bands calculated from the $13$ MLWFs, despite the entanglement with the Ti-$e_{\rm g}$ bands. In Fig.~\ref{fig:OV01_dft}b, where the projected density of states of the MLWFs is shown, we can see that the Ti-\tgg bands are essentially described by the set of $12$ Ti-centered Wannier functions, with both inequivalent Ti presenting very similar features. We can represent essentially the whole weight of the vacancy Bloch state with a single Wannier function (orange line Fig.~\ref{fig:OV01_dft}b). Interesting and crucial to our approach is the fact that in the real-space representation of the corresponding Wannier function (Fig.~\ref{fig:OV01_dft}c), the spheroidal charge is centered approximately where the missing oxygen would be in the pure case, with tails on the surrounding atoms. For the rest of this work, we will focus on the case of the axial vacancy since our conclusions regarding the stability of the Mott insulating state are the same for both configurations. 

\begin{figure*}
	\includegraphics[width=0.9\textwidth]{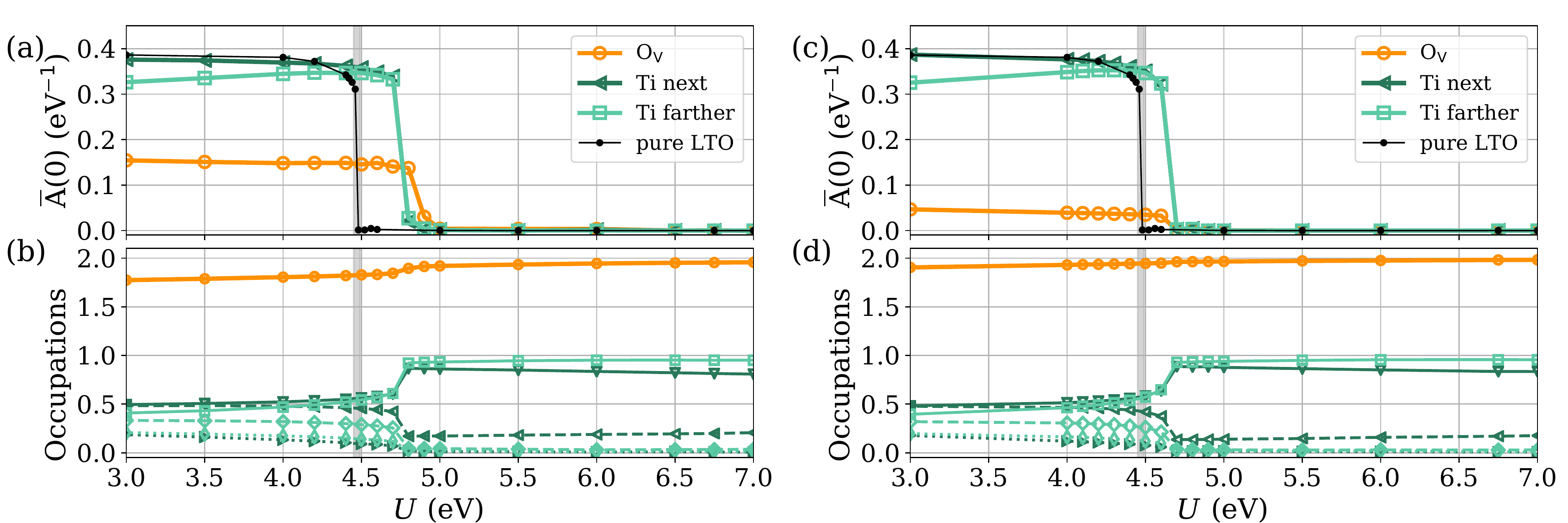}
	\caption{(Color online) Spectral weight at the Fermi level, $\bar{A}(0)$ (a, c) and orbital occupations (b, d) for the two inequivalent Ti atoms (dark and light green lines) and for the vacancy state (orange line), obtained from DMFT. The left column (a, b) represents the case for an ``uncorrelated'' vacancy site, with $U$ ($x$ axis) applied only to the Ti sites, while $U($\ov$)=0$. In the right column (c, d), $U$ is the same for every site, $U($Ti$)=U($\ov$)$. Solid black lines (a, c) show $\bar{A}(0)$ for pure \LTO, and the gray vertical lines its $U_{\rm{MIT}}$.}
	\label{fig:OV_occ}
\end{figure*}

\begin{figure*}
	\includegraphics[width=0.8\textwidth]{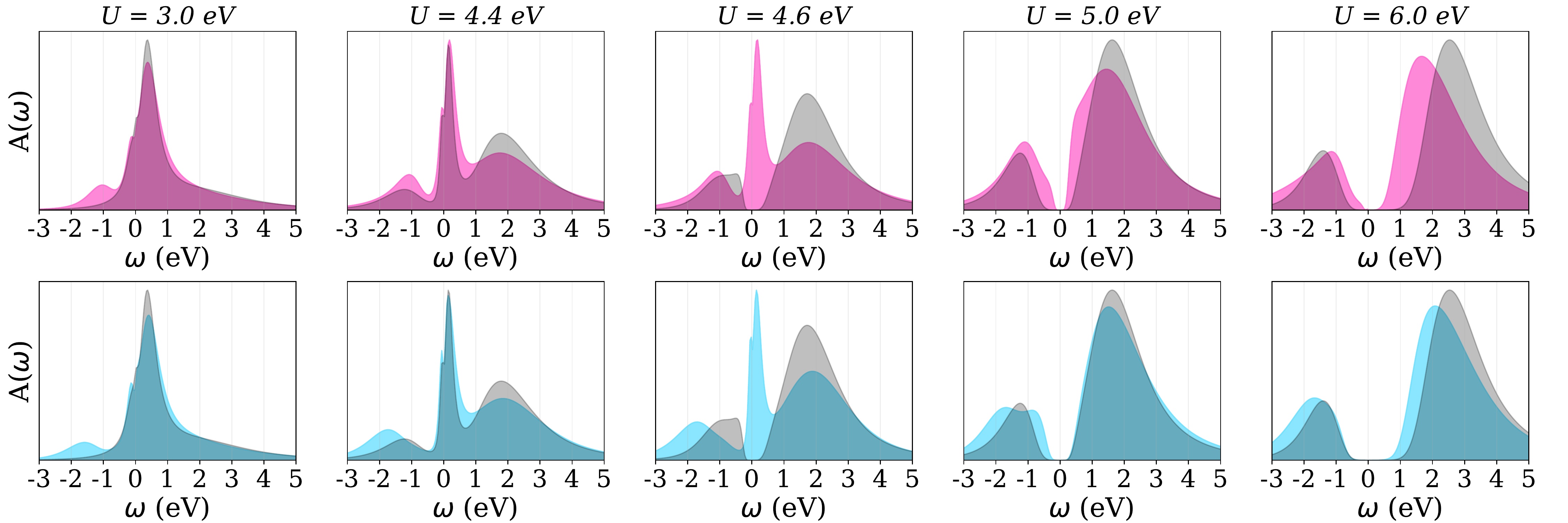}
	\caption{(Color online) Comparison of the total spectral functions $A(\omega)$ between the \LTOovv system (colored area) and pure defect-free \LTOO (gray area) obtained within DMFT for different values of $U$. The upper row (magenta) shows the results for the case where $U($\ov$)$ is zero. In the lower row (blue) all sites have the same $U$, $U($Ti$)=U($\ov$)$.
	 $A(\omega)$ plots have been normalized to the total number of states within the corresponding basis.}
	\label{fig:OV_aws}
\end{figure*}

Next, we perform paramagnetic DMFT calculations based on the fixed input electronic structure of the Hamiltonian expressed in the MLWF-basis.
As stated earlier, an effective 3-orbital impurity problem is solved for each inequivalent Ti site plus an additional 1-orbital impurity problem for the O$_{\rm V}$-centered MLWF. Initially, we treat the vacancy as ``uncorrelated'', i.e., we set $U($\ov$)$ to zero, while keeping $U({\rm Ti})$ non-zero. For simplicity, we use the same value of $U$ on each Ti site. In Fig. \ref{fig:OV_occ}a and \ref{fig:OV_occ}b, the calculated spectral weight at the Fermi level and the corresponding orbital occupations are shown as a function of $U({\rm Ti})$ for all inequivalent sites of the system, i.e., two Ti sites and one \ovv site\cite{nota_occupations}. Our DMFT calculations for vacancy-free \LTOO (solid black line in Fig.~\ref{fig:OV_occ}a) show a clear MIT at $U_{\rm{MIT}} = 4.5$ eV, indicated by a sudden drop in the spectral weight to $\bar{A}(0) \approx 0$, and thus a separation between a metallic regime (for $U<U_{\rm MIT}$) and a Mott-insulating regime (for $U>U_{\rm MIT}$). This is consistent with previous DFT+DMFT studies~\cite{Pavarini,PhysRevB.89.161109}, where a value of $U \sim 5.0$ eV, slightly above $U_{\rm{MIT}}$, is often used to give a realistic description of \LTO.

The changes in \LTOovv are subtle with respect to the defect-free material. Importantly, the system still undergoes a MIT with $U_{\rm{MIT}} = 4.8$ eV, a slightly higher value (by ${\sim}0.3$ eV) than the stoichiometric case. For all sites (including the vacancy site) the spectral weight $\bar{A}(0)$ drops to zero at essentially the same value of $U_{\rm{MIT}}$. In the metallic state, we find all three Ti-\tgg orbitals to be fractionally occupied, while in the insulating state the system exhibits a strong orbital polarization with one $t_{2g}$ orbital nearly completely filled, and the other two nearly empty, again consistent with results for the stoichiometric system~\cite{Pavarini,PhysRevB.89.161109}. However, the orbital polarization in the Mott-insulating state is reduced for the Ti atom next to the vacancy, for which the total charge of one electron is split between two \tgg orbitals with occupations $0.8$ and $0.2$ ($U=7.0$ eV). For the whole range of $U$, the vacancy site stays close to doubly occupied, in particular in the insulating regime for $U>U_{\rm MIT}$, with a maximum charge transfer of ${\sim}0.22$ e$^-$ to the neighboring Ti (for $U=3.0$ eV). This charge transfer gradually decreases until it nearly vanishes at $U_{\rm MIT}$. 

Finally, we perform calculations with a non-zero $U$ on the vacancy site, treating explicitly its correlations, setting $U($\ov$)=U($Ti$)$ for simplicity.
Given the larger spread shown by the calculated \ov Wannier function compared to those describing the Ti-\tgg states, and the inverse relation between the localization of a Wannier function and its on-site Coulomb energy repulsion, this choice of $U($\ov$)$ can be viewed as an upper limit. In Fig.~\ref{fig:OV_occ}c and \ref{fig:OV_occ}d we see that the changes introduced by the defect are smaller than for the case with $U($\ov$)=0$. The critical $U$ value for the MIT is now $U_{\rm{MIT}} = 4.7$ eV, closer to the value for the defect-free system. In addition, the amount of charge donated to the neighboring Ti from \ovv is further reduced, becoming almost negligible even below $U_{\rm{MIT}}$.

These differences are also reflected in the total spectral functions shown in Fig.~\ref{fig:OV_aws}. For a small $U=3$~eV, the main difference between the defect-free case and the two cases with O$_{\rm V}$ is the presence of the additional vacancy band centered around $\sim$$-1.5$~eV. Approaching $U_{\rm{MIT}}$, a three-peaked structure, the well-known hallmark of a strongly correlated metal, emerges for all three scenarios.
Similarly to the case of the correlated metal SrVO$_3$ discussed in Ref.~\onlinecite{PhysRevB.94.241110}, the lower Hubbard band emerges at approximately the energy where the vacancy band is located, making it experimentally hard to distinguish between the two.
The slight increase of $U_{\rm MIT}$ due to the presence of the vacancy is apparent from the $U=4.6$ eV panel where pure \LTOO is already gapped, while both \LTOovv cases still show a prominent quasiparticle peak. The next panel to the right, $U=5.0$ eV, shows that the case of \LTOovv with $U($\ov$)=U({\rm{Ti}})$ (blue data) exhibits a gap only slightly smaller than defect-free \LTO, whereas in the case with $U($\ov$)=0$ (magenta data) the gap is still minimal. Deeper within the insulating phase, for $U=6.0$ eV, the gap still remains smaller for $U($\ov$)=0$, compared to the defect-free material and the case with nonzero $U($\ov$)$.

\begin{figure}
	\includegraphics[width=0.4\textwidth]{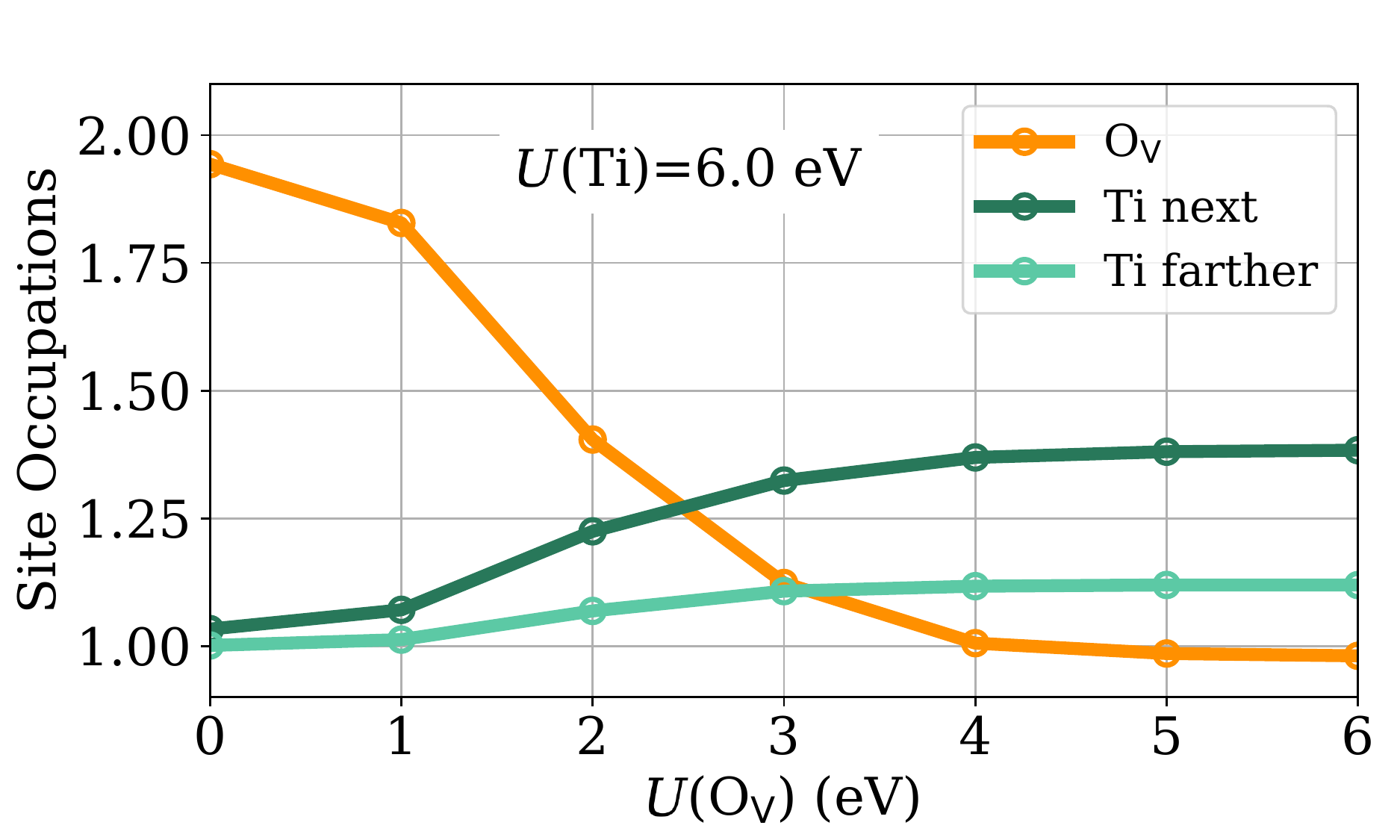}
	\caption{(Color online) DMFT site occupations as a function of $U($\ov$)$ with the double-counting term on the vacancy site set to zero. $U($Ti$)$ is kept fixed at $6.0$ eV. Without the double-counting term shifting the vacancy band down in energy, the vacancy site evolves towards a half-filled state, the Ti sites accept the excess electron, and the system becomes metallic.}
	\label{fig:DCoffOV}
\end{figure}

From the localization of the corresponding Wannier functions,
we expect that a realistic value for $U$ on the vacancy is somewhere
in between the two cases studied here, $0 < U($\ov$) <
U(\rm{Ti})$. We therefore also perform DMFT calculations with
$U($Ti$)$ kept constant while $U($\ov$)$ is varied. We find
that the resulting occupations of all sites are largely independent
of $U($\ov$)$. Furthermore, the system remains either metallic or
insulating, depending on whether $U(\rm{Ti})$ is below or above
$U=4.6$ eV. Only for  $U($Ti$)=4.6$ eV, where the system is
still metallic for both $U($\ov$)=0$ and $U($\ov$)=U(\rm{Ti})$, see
Figs.~\ref{fig:OV_occ} and \ref{fig:OV_aws}, does increasing the Hubbard
$U$ on \ovv trigger a transition to the insulating state, albeit for a
rather high $U($\ov$){\sim} 7.0$ eV. Thus, we find that for $U($Ti$) > 4.7$ eV the system remains insulating for any $U($\ov$)$, while in the narrow range of $4.5$ eV $< U($Ti$) \leq 4.7$ eV the incorporation of oxygen vacancies can potentially destroy the Mott insulating character. However, even in this case, the vacancy band stays essentially fully occupied and the MIT is rather caused by subtle changes in the electronic structure. We emphasize that such a scenario would be completely different from doping of the insulating state by the electrons left behind by the oxygen vacancy, which does not occur for any reasonable value of $U$.

In order to understand why increasing the Coulomb repulsion on the vacancy site does not lead to a depletion of the corresponding states and a charge transfer into the Ti-\tg, one has to consider the effect of the double-counting correction in our DFT+DMFT calculations. This correction
attempts to subtract the effect of the electron-electron interaction
within the correlated subspace that is included both on the DFT and
the DMFT level, in order to avoid double counting. In practice, it
enters the calculation as a local potential shift that depends
linearly on both $U$ and the local occupation, and is
applied to all Ti sites as well as to the O$_{\rm V}$. Due to the
higher occupation of the O$_{\rm V}$ compared to the Ti sites, the
double-counting shift is larger for the former (for similar $U$
values), i.e., the double-counting term shifts the vacancy states further
down in energy, thereby reinforcing the double occupation of these
states and preventing any charge transfer into the higher-lying Ti
bands.

If the double-counting term for the vacancy site is
nullified within the calculation (Fig.~\ref{fig:DCoffOV}), the
\ovv site is indeed depleted, reaching the half-filled state for
$U($\ov$) \gtrsim 3$ eV (even for $U($Ti$)=6.0$~eV, deep inside the insulating
regime for the pure system). The system then stays in
a metallic regime even for values of $U($Ti$)$ where pure \LTOO or
\LTO$_{-\delta}$ with full double counting already behaves as a Mott
insulator. We note that, while there is uncertainty regarding the
most appropriate form of the double-counting
correction \cite{Karolak2010}, completely neglecting the double counting
is clearly unphysical, and is done here only for test
purposes.  We find that, scaling down the value of
the double-counting term applied to \ovv to ${\sim}75\%$ of its original value still results
in a doubly occupied vacancy and a stable Mott-insulating state. 
We therefore conclude that, in spite of any uncertainties
regarding the exact value of the double-counting correction, our
DFT+DMFT calculations predict a very weak effect of the vacancy states
on the Mott-insulating character of \LTO.
The charge released by the \ovv remains mainly on the vacancy sites and does
not change the filling of the Ti-$t_{2g}$ bands, thus leaving the Mott insulator
essentially unperturbed with respect to the situation in pure \LTO.


To conclude, we have performed DFT+DMFT calculations for \ov-defective \LTOO to investigate the effect of such vacancies on the Mott-insulating state. We find that the presence of the vacancy creates new states at energies slightly below the partially occupied Ti-$t_{2g}$ bands. Provided that the Coulomb repulsion is not extremely close to the critical value for the MIT within the pure compound, these defect states remain doubly-occupied, and therefore will not change the filling of the Ti bands and thus affect the Mott-insulating character of \LTO. In spite of the relatively small gap of \LTO, reflecting its close vicinity to the MIT, its Mott insulating character is surprisingly robust against the incorporation of oxygen vacancies, that do not dope the system for any value of the interaction strength.

The explicit treatment of the vacancy state developed in this work
provides an efficient and physically transparent way to study electronic correlations in defective systems,
requiring no prior assumptions about the nature of the
defect states. This complements other approaches of defect characterization~\cite{PhysRevLett.111.217601}, and will hopefully motivate studies in other Mott materials with different kinds of point defects.



This work was supported by the Swiss National Science Foundation through NCCR-MARVEL.
Calculations have been performed on the cluster ``M\"onch'' and ``Piz Daint'', both
hosted by the Swiss National Supercomputing Centre, and the ``Euler'' clusters of ETH Zurich. \\

\bibliography{references}

\begin{thebibliography}{43}%
\makeatletter
\providecommand \@ifxundefined [1]{%
 \@ifx{#1\undefined}
}%
\providecommand \@ifnum [1]{%
 \ifnum #1\expandafter \@firstoftwo
 \else \expandafter \@secondoftwo
 \fi
}%
\providecommand \@ifx [1]{%
 \ifx #1\expandafter \@firstoftwo
 \else \expandafter \@secondoftwo
 \fi
}%
\providecommand \natexlab [1]{#1}%
\providecommand \enquote  [1]{``#1''}%
\providecommand \bibnamefont  [1]{#1}%
\providecommand \bibfnamefont [1]{#1}%
\providecommand \citenamefont [1]{#1}%
\providecommand \href@noop [0]{\@secondoftwo}%
\providecommand \href [0]{\begingroup \@sanitize@url \@href}%
\providecommand \@href[1]{\@@startlink{#1}\@@href}%
\providecommand \@@href[1]{\endgroup#1\@@endlink}%
\providecommand \@sanitize@url [0]{\catcode `\\12\catcode `\$12\catcode
  `\&12\catcode `\#12\catcode `\^12\catcode `\_12\catcode `\%12\relax}%
\providecommand \@@startlink[1]{}%
\providecommand \@@endlink[0]{}%
\providecommand \url  [0]{\begingroup\@sanitize@url \@url }%
\providecommand \@url [1]{\endgroup\@href {#1}{\urlprefix }}%
\providecommand \urlprefix  [0]{URL }%
\providecommand \Eprint [0]{\href }%
\providecommand \doibase [0]{http://dx.doi.org/}%
\providecommand \selectlanguage [0]{\@gobble}%
\providecommand \bibinfo  [0]{\@secondoftwo}%
\providecommand \bibfield  [0]{\@secondoftwo}%
\providecommand \translation [1]{[#1]}%
\providecommand \BibitemOpen [0]{}%
\providecommand \bibitemStop [0]{}%
\providecommand \bibitemNoStop [0]{.\EOS\space}%
\providecommand \EOS [0]{\spacefactor3000\relax}%
\providecommand \BibitemShut  [1]{\csname bibitem#1\endcsname}%
\let\auto@bib@innerbib\@empty
\bibitem [{tag()}]{nota_OVnotation}%
  \BibitemOpen
  \bibinfo {note} {{I}n this work, O$_{\rm
  V}$ will always refer to the {\it neutral} O$_{\rm V}$, ${\rm V}_{\,{\rm
  O}}^{\bullet\bullet}$ in the Kr\"oger-Vink notation.}\BibitemShut {Stop}%
\bibitem [{\citenamefont {Cava}\ \emph {et~al.}(1987)\citenamefont {Cava},
  \citenamefont {Batlogg}, \citenamefont {Chen}, \citenamefont {Rietman},
  \citenamefont {Zahurak},\ and\ \citenamefont {Werder}}]{Cava1987}%
  \BibitemOpen
  \bibfield  {author} {\bibinfo {author} {\bibfnamefont {R.~J.}\ \bibnamefont
  {Cava}}, \bibinfo {author} {\bibfnamefont {B.}~\bibnamefont {Batlogg}},
  \bibinfo {author} {\bibfnamefont {C.~H.}\ \bibnamefont {Chen}}, \bibinfo
  {author} {\bibfnamefont {E.~A.}\ \bibnamefont {Rietman}}, \bibinfo {author}
  {\bibfnamefont {S.~M.}\ \bibnamefont {Zahurak}}, \ and\ \bibinfo {author}
  {\bibfnamefont {D.}~\bibnamefont {Werder}},\ }\href
  {http://dx.doi.org/10.1038/329423a0} {\bibfield  {journal} {\bibinfo
  {journal} {Nature}\ }\textbf {\bibinfo {volume} {329}},\ \bibinfo {pages}
  {423} (\bibinfo {year} {1987})}\BibitemShut {NoStop}%
\bibitem [{\citenamefont {Kalabukhov}\ \emph {et~al.}(2007)\citenamefont
  {Kalabukhov}, \citenamefont {Gunnarsson}, \citenamefont {B\"orjesson},
  \citenamefont {Olsson}, \citenamefont {Claeson},\ and\ \citenamefont
  {Winkler}}]{PhysRevB.75.121404}%
  \BibitemOpen
  \bibfield  {author} {\bibinfo {author} {\bibfnamefont {A.}~\bibnamefont
  {Kalabukhov}}, \bibinfo {author} {\bibfnamefont {R.}~\bibnamefont
  {Gunnarsson}}, \bibinfo {author} {\bibfnamefont {J.}~\bibnamefont
  {B\"orjesson}}, \bibinfo {author} {\bibfnamefont {E.}~\bibnamefont {Olsson}},
  \bibinfo {author} {\bibfnamefont {T.}~\bibnamefont {Claeson}}, \ and\
  \bibinfo {author} {\bibfnamefont {D.}~\bibnamefont {Winkler}},\ }\href
  {\doibase 10.1103/PhysRevB.75.121404} {\bibfield  {journal} {\bibinfo
  {journal} {Phys. Rev. B}\ }\textbf {\bibinfo {volume} {75}},\ \bibinfo
  {pages} {121404} (\bibinfo {year} {2007})}\BibitemShut {NoStop}%
\bibitem [{\citenamefont {Kormondy}\ \emph {et~al.}(2018)\citenamefont
  {Kormondy}, \citenamefont {Gao}, \citenamefont {Li}, \citenamefont {Lu},
  \citenamefont {Posadas}, \citenamefont {Shen}, \citenamefont {Tsoi},
  \citenamefont {McCartney}, \citenamefont {Smith}, \citenamefont {Zhou},
  \citenamefont {Lev}, \citenamefont {Husanu}, \citenamefont {Strocov},\ and\
  \citenamefont {Demkov}}]{Kormondy2018}%
  \BibitemOpen
  \bibfield  {author} {\bibinfo {author} {\bibfnamefont {K.~J.}\ \bibnamefont
  {Kormondy}}, \bibinfo {author} {\bibfnamefont {L.}~\bibnamefont {Gao}},
  \bibinfo {author} {\bibfnamefont {X.}~\bibnamefont {Li}}, \bibinfo {author}
  {\bibfnamefont {S.}~\bibnamefont {Lu}}, \bibinfo {author} {\bibfnamefont
  {A.~B.}\ \bibnamefont {Posadas}}, \bibinfo {author} {\bibfnamefont
  {S.}~\bibnamefont {Shen}}, \bibinfo {author} {\bibfnamefont {M.}~\bibnamefont
  {Tsoi}}, \bibinfo {author} {\bibfnamefont {M.~R.}\ \bibnamefont {McCartney}},
  \bibinfo {author} {\bibfnamefont {D.~J.}\ \bibnamefont {Smith}}, \bibinfo
  {author} {\bibfnamefont {J.}~\bibnamefont {Zhou}}, \bibinfo {author}
  {\bibfnamefont {L.~L.}\ \bibnamefont {Lev}}, \bibinfo {author} {\bibfnamefont
  {M.-A.}\ \bibnamefont {Husanu}}, \bibinfo {author} {\bibfnamefont {V.~N.}\
  \bibnamefont {Strocov}}, \ and\ \bibinfo {author} {\bibfnamefont {A.~A.}\
  \bibnamefont {Demkov}},\ }\href {\doibase 10.1038/s41598-018-26017-z}
  {\bibfield  {journal} {\bibinfo  {journal} {Scientific Reports}\ }\textbf
  {\bibinfo {volume} {8}},\ \bibinfo {pages} {7721} (\bibinfo {year}
  {2018})}\BibitemShut {NoStop}%
\bibitem [{\citenamefont {Hwang}(2005)}]{Hwang2005}%
  \BibitemOpen
  \bibfield  {author} {\bibinfo {author} {\bibfnamefont {H.~Y.}\ \bibnamefont
  {Hwang}},\ }\href {http://dx.doi.org/10.1038/nmat1515} {\bibfield  {journal}
  {\bibinfo  {journal} {Nature Materials}\ }\textbf {\bibinfo {volume} {4}},\
  \bibinfo {pages} {803} (\bibinfo {year} {2005})}\BibitemShut {NoStop}%
\bibitem [{\citenamefont {Crespillo}\ \emph {et~al.}(2017)\citenamefont
  {Crespillo}, \citenamefont {Graham}, \citenamefont {Agull\'o-L\'opez},
  \citenamefont {Zhang},\ and\ \citenamefont {Weber}}]{Crespillo2017}%
  \BibitemOpen
  \bibfield  {author} {\bibinfo {author} {\bibfnamefont {M.~L.}\ \bibnamefont
  {Crespillo}}, \bibinfo {author} {\bibfnamefont {J.~T.}\ \bibnamefont
  {Graham}}, \bibinfo {author} {\bibfnamefont {F.}~\bibnamefont
  {Agull\'o-L\'opez}}, \bibinfo {author} {\bibfnamefont {Y.}~\bibnamefont
  {Zhang}}, \ and\ \bibinfo {author} {\bibfnamefont {W.~J.}\ \bibnamefont
  {Weber}},\ }\href {http://stacks.iop.org/0022-3727/50/i=15/a=155303}
  {\bibfield  {journal} {\bibinfo  {journal} {Journal of Physics D: Applied
  Physics}\ }\textbf {\bibinfo {volume} {50}},\ \bibinfo {pages} {155303}
  (\bibinfo {year} {2017})}\BibitemShut {NoStop}%
\bibitem [{\citenamefont {Ganduglia-Pirovano}\ \emph
  {et~al.}(2007)\citenamefont {Ganduglia-Pirovano}, \citenamefont {Hofmann},\
  and\ \citenamefont {Sauer}}]{ganduglia2007oxygen}%
  \BibitemOpen
  \bibfield  {author} {\bibinfo {author} {\bibfnamefont {M.~V.}\ \bibnamefont
  {Ganduglia-Pirovano}}, \bibinfo {author} {\bibfnamefont {A.}~\bibnamefont
  {Hofmann}}, \ and\ \bibinfo {author} {\bibfnamefont {J.}~\bibnamefont
  {Sauer}},\ }\href {\doibase https://doi.org/10.1016/j.surfrep.2007.03.002}
  {\bibfield  {journal} {\bibinfo  {journal} {Surface Science Reports}\
  }\textbf {\bibinfo {volume} {62}},\ \bibinfo {pages} {219 } (\bibinfo {year}
  {2007})}\BibitemShut {NoStop}%
\bibitem [{\citenamefont {Eckstein}(2007)}]{Eckstein2007}%
  \BibitemOpen
  \bibfield  {author} {\bibinfo {author} {\bibfnamefont {J.~N.}\ \bibnamefont
  {Eckstein}},\ }\href {http://dx.doi.org/10.1038/nmat1944} {\bibfield
  {journal} {\bibinfo  {journal} {Nature Materials}\ }\textbf {\bibinfo
  {volume} {6}},\ \bibinfo {pages} {473} (\bibinfo {year} {2007})}\BibitemShut
  {NoStop}%
\bibitem [{\citenamefont {Backes}\ \emph {et~al.}(2016)\citenamefont {Backes},
  \citenamefont {R\"odel}, \citenamefont {Fortuna}, \citenamefont
  {Frantzeskakis}, \citenamefont {Le~F\`evre}, \citenamefont {Bertran},
  \citenamefont {Kobayashi}, \citenamefont {Yukawa}, \citenamefont
  {Mitsuhashi}, \citenamefont {Kitamura}, \citenamefont {Horiba}, \citenamefont
  {Kumigashira}, \citenamefont {Saint-Martin}, \citenamefont {Fouchet},
  \citenamefont {Berini}, \citenamefont {Dumont}, \citenamefont {Kim},
  \citenamefont {Lechermann}, \citenamefont {Jeschke}, \citenamefont
  {Rozenberg}, \citenamefont {Valent\'{\i}},\ and\ \citenamefont
  {Santander-Syro}}]{PhysRevB.94.241110}%
  \BibitemOpen
  \bibfield  {author} {\bibinfo {author} {\bibfnamefont {S.}~\bibnamefont
  {Backes}}, \bibinfo {author} {\bibfnamefont {T.~C.}\ \bibnamefont {R\"odel}},
  \bibinfo {author} {\bibfnamefont {F.}~\bibnamefont {Fortuna}}, \bibinfo
  {author} {\bibfnamefont {E.}~\bibnamefont {Frantzeskakis}}, \bibinfo {author}
  {\bibfnamefont {P.}~\bibnamefont {Le~F\`evre}}, \bibinfo {author}
  {\bibfnamefont {F.}~\bibnamefont {Bertran}}, \bibinfo {author} {\bibfnamefont
  {M.}~\bibnamefont {Kobayashi}}, \bibinfo {author} {\bibfnamefont
  {R.}~\bibnamefont {Yukawa}}, \bibinfo {author} {\bibfnamefont
  {T.}~\bibnamefont {Mitsuhashi}}, \bibinfo {author} {\bibfnamefont
  {M.}~\bibnamefont {Kitamura}}, \bibinfo {author} {\bibfnamefont
  {K.}~\bibnamefont {Horiba}}, \bibinfo {author} {\bibfnamefont
  {H.}~\bibnamefont {Kumigashira}}, \bibinfo {author} {\bibfnamefont
  {R.}~\bibnamefont {Saint-Martin}}, \bibinfo {author} {\bibfnamefont
  {A.}~\bibnamefont {Fouchet}}, \bibinfo {author} {\bibfnamefont
  {B.}~\bibnamefont {Berini}}, \bibinfo {author} {\bibfnamefont
  {Y.}~\bibnamefont {Dumont}}, \bibinfo {author} {\bibfnamefont {A.~J.}\
  \bibnamefont {Kim}}, \bibinfo {author} {\bibfnamefont {F.}~\bibnamefont
  {Lechermann}}, \bibinfo {author} {\bibfnamefont {H.~O.}\ \bibnamefont
  {Jeschke}}, \bibinfo {author} {\bibfnamefont {M.~J.}\ \bibnamefont
  {Rozenberg}}, \bibinfo {author} {\bibfnamefont {R.}~\bibnamefont
  {Valent\'{\i}}}, \ and\ \bibinfo {author} {\bibfnamefont {A.~F.}\
  \bibnamefont {Santander-Syro}},\ }\href {\doibase 10.1103/PhysRevB.94.241110}
  {\bibfield  {journal} {\bibinfo  {journal} {Phys. Rev. B}\ }\textbf {\bibinfo
  {volume} {94}},\ \bibinfo {pages} {241110} (\bibinfo {year}
  {2016})}\BibitemShut {NoStop}%
\bibitem [{\citenamefont {Janotti}\ \emph {et~al.}(2014)\citenamefont
  {Janotti}, \citenamefont {Varley}, \citenamefont {Choi},\ and\ \citenamefont
  {Van~de Walle}}]{PhysRevB.90.085202}%
  \BibitemOpen
  \bibfield  {author} {\bibinfo {author} {\bibfnamefont {A.}~\bibnamefont
  {Janotti}}, \bibinfo {author} {\bibfnamefont {J.~B.}\ \bibnamefont {Varley}},
  \bibinfo {author} {\bibfnamefont {M.}~\bibnamefont {Choi}}, \ and\ \bibinfo
  {author} {\bibfnamefont {C.~G.}\ \bibnamefont {Van~de Walle}},\ }\href
  {\doibase 10.1103/PhysRevB.90.085202} {\bibfield  {journal} {\bibinfo
  {journal} {Phys. Rev. B}\ }\textbf {\bibinfo {volume} {90}},\ \bibinfo
  {pages} {085202} (\bibinfo {year} {2014})}\BibitemShut {NoStop}%
\bibitem [{\citenamefont {Lin}\ and\ \citenamefont
  {Demkov}(2013)}]{PhysRevLett.111.217601}%
  \BibitemOpen
  \bibfield  {author} {\bibinfo {author} {\bibfnamefont {C.}~\bibnamefont
  {Lin}}\ and\ \bibinfo {author} {\bibfnamefont {A.~A.}\ \bibnamefont
  {Demkov}},\ }\href {\doibase 10.1103/PhysRevLett.111.217601} {\bibfield
  {journal} {\bibinfo  {journal} {Phys. Rev. Lett.}\ }\textbf {\bibinfo
  {volume} {111}},\ \bibinfo {pages} {217601} (\bibinfo {year}
  {2013})}\BibitemShut {NoStop}%
\bibitem [{\citenamefont {Lechermann}\ \emph {et~al.}(2016)\citenamefont
  {Lechermann}, \citenamefont {Jeschke}, \citenamefont {Kim}, \citenamefont
  {Backes},\ and\ \citenamefont {Valent\'{\i}}}]{PhysRevB.93.121103}%
  \BibitemOpen
  \bibfield  {author} {\bibinfo {author} {\bibfnamefont {F.}~\bibnamefont
  {Lechermann}}, \bibinfo {author} {\bibfnamefont {H.~O.}\ \bibnamefont
  {Jeschke}}, \bibinfo {author} {\bibfnamefont {A.~J.}\ \bibnamefont {Kim}},
  \bibinfo {author} {\bibfnamefont {S.}~\bibnamefont {Backes}}, \ and\ \bibinfo
  {author} {\bibfnamefont {R.}~\bibnamefont {Valent\'{\i}}},\ }\href {\doibase
  10.1103/PhysRevB.93.121103} {\bibfield  {journal} {\bibinfo  {journal} {Phys.
  Rev. B}\ }\textbf {\bibinfo {volume} {93}},\ \bibinfo {pages} {121103}
  (\bibinfo {year} {2016})}\BibitemShut {NoStop}%
\bibitem [{\citenamefont {Imada}\ \emph {et~al.}(1998)\citenamefont {Imada},
  \citenamefont {Fujimori},\ and\ \citenamefont {Tokura}}]{RevModPhys.70.1039}%
  \BibitemOpen
  \bibfield  {author} {\bibinfo {author} {\bibfnamefont {M.}~\bibnamefont
  {Imada}}, \bibinfo {author} {\bibfnamefont {A.}~\bibnamefont {Fujimori}}, \
  and\ \bibinfo {author} {\bibfnamefont {Y.}~\bibnamefont {Tokura}},\ }\href
  {\doibase 10.1103/RevModPhys.70.1039} {\bibfield  {journal} {\bibinfo
  {journal} {Rev. Mod. Phys.}\ }\textbf {\bibinfo {volume} {70}},\ \bibinfo
  {pages} {1039} (\bibinfo {year} {1998})}\BibitemShut {NoStop}%
\bibitem [{\citenamefont {Mannhart}\ and\ \citenamefont
  {Haensch}(2012)}]{Mannhart2012}%
  \BibitemOpen
  \bibfield  {author} {\bibinfo {author} {\bibfnamefont {J.}~\bibnamefont
  {Mannhart}}\ and\ \bibinfo {author} {\bibfnamefont {W.}~\bibnamefont
  {Haensch}},\ }\href {https://doi.org/10.1038/487436a} {\bibfield  {journal}
  {\bibinfo  {journal} {Nature}\ }\textbf {\bibinfo {volume} {487}},\ \bibinfo
  {pages} {436} (\bibinfo {year} {2012})}\BibitemShut {NoStop}%
\bibitem [{\citenamefont {Inoue}\ and\ \citenamefont
  {Rozenberg}(2008)}]{Inoue2008}%
  \BibitemOpen
  \bibfield  {author} {\bibinfo {author} {\bibfnamefont {I.~H.}\ \bibnamefont
  {Inoue}}\ and\ \bibinfo {author} {\bibfnamefont {M.~J.}\ \bibnamefont
  {Rozenberg}},\ }\href {\doibase 10.1002/adfm.200800558} {\bibfield  {journal}
  {\bibinfo  {journal} {Adv. Funct. Mater.}\ }\textbf {\bibinfo {volume}
  {18}},\ \bibinfo {pages} {2289} (\bibinfo {year} {2008})}\BibitemShut
  {NoStop}%
\bibitem [{\citenamefont {Aschauer}\ \emph {et~al.}(2013)\citenamefont
  {Aschauer}, \citenamefont {Pfenninger}, \citenamefont {Selbach},
  \citenamefont {Grande},\ and\ \citenamefont {Spaldin}}]{PhysRevB.88.054111}%
  \BibitemOpen
  \bibfield  {author} {\bibinfo {author} {\bibfnamefont {U.}~\bibnamefont
  {Aschauer}}, \bibinfo {author} {\bibfnamefont {R.}~\bibnamefont
  {Pfenninger}}, \bibinfo {author} {\bibfnamefont {S.~M.}\ \bibnamefont
  {Selbach}}, \bibinfo {author} {\bibfnamefont {T.}~\bibnamefont {Grande}}, \
  and\ \bibinfo {author} {\bibfnamefont {N.~A.}\ \bibnamefont {Spaldin}},\
  }\href {\doibase 10.1103/PhysRevB.88.054111} {\bibfield  {journal} {\bibinfo
  {journal} {Phys. Rev. B}\ }\textbf {\bibinfo {volume} {88}},\ \bibinfo
  {pages} {054111} (\bibinfo {year} {2013})}\BibitemShut {NoStop}%
\bibitem [{\citenamefont {Amaricci}\ \emph {et~al.}(2017)\citenamefont
  {Amaricci}, \citenamefont {de’ Medici},\ and\ \citenamefont
  {Capone}}]{0295-5075-118-1-17004}%
  \BibitemOpen
  \bibfield  {author} {\bibinfo {author} {\bibfnamefont {A.}~\bibnamefont
  {Amaricci}}, \bibinfo {author} {\bibfnamefont {L.}~\bibnamefont {de’
  Medici}}, \ and\ \bibinfo {author} {\bibfnamefont {M.}~\bibnamefont
  {Capone}},\ }\href {http://stacks.iop.org/0295-5075/118/i=1/a=17004}
  {\bibfield  {journal} {\bibinfo  {journal} {EPL (Europhysics Letters)}\
  }\textbf {\bibinfo {volume} {118}},\ \bibinfo {pages} {17004} (\bibinfo
  {year} {2017})}\BibitemShut {NoStop}%
\bibitem [{\citenamefont {Werner}\ \emph {et~al.}(2009)\citenamefont {Werner},
  \citenamefont {Gull},\ and\ \citenamefont {Millis}}]{PhysRevB.79.115119}%
  \BibitemOpen
  \bibfield  {author} {\bibinfo {author} {\bibfnamefont {P.}~\bibnamefont
  {Werner}}, \bibinfo {author} {\bibfnamefont {E.}~\bibnamefont {Gull}}, \ and\
  \bibinfo {author} {\bibfnamefont {A.~J.}\ \bibnamefont {Millis}},\ }\href
  {\doibase 10.1103/PhysRevB.79.115119} {\bibfield  {journal} {\bibinfo
  {journal} {Phys. Rev. B}\ }\textbf {\bibinfo {volume} {79}},\ \bibinfo
  {pages} {115119} (\bibinfo {year} {2009})}\BibitemShut {NoStop}%
\bibitem [{\citenamefont {Georges}\ \emph {et~al.}(1996)\citenamefont
  {Georges}, \citenamefont {Kotliar}, \citenamefont {Krauth},\ and\
  \citenamefont {Rozenberg}}]{georges1996dynamical}%
  \BibitemOpen
  \bibfield  {author} {\bibinfo {author} {\bibfnamefont {A.}~\bibnamefont
  {Georges}}, \bibinfo {author} {\bibfnamefont {G.}~\bibnamefont {Kotliar}},
  \bibinfo {author} {\bibfnamefont {W.}~\bibnamefont {Krauth}}, \ and\ \bibinfo
  {author} {\bibfnamefont {M.~J.}\ \bibnamefont {Rozenberg}},\ }\href {\doibase
  10.1103/RevModPhys.68.13} {\bibfield  {journal} {\bibinfo  {journal} {Rev.
  Mod. Phys.}\ }\textbf {\bibinfo {volume} {68}},\ \bibinfo {pages} {13}
  (\bibinfo {year} {1996})}\BibitemShut {NoStop}%
\bibitem [{\citenamefont {Anisimov}\ \emph {et~al.}(1997)\citenamefont
  {Anisimov}, \citenamefont {Poteryaev}, \citenamefont {Korotin}, \citenamefont
  {Anokhin},\ and\ \citenamefont {Kotliar}}]{anisimov1997first}%
  \BibitemOpen
  \bibfield  {author} {\bibinfo {author} {\bibfnamefont {V.~I.}\ \bibnamefont
  {Anisimov}}, \bibinfo {author} {\bibfnamefont {A.~I.}\ \bibnamefont
  {Poteryaev}}, \bibinfo {author} {\bibfnamefont {M.~A.}\ \bibnamefont
  {Korotin}}, \bibinfo {author} {\bibfnamefont {A.~O.}\ \bibnamefont
  {Anokhin}}, \ and\ \bibinfo {author} {\bibfnamefont {G.}~\bibnamefont
  {Kotliar}},\ }\href {http://stacks.iop.org/0953-8984/9/i=35/a=010} {\bibfield
   {journal} {\bibinfo  {journal} {Journal of Physics: Condensed Matter}\
  }\textbf {\bibinfo {volume} {9}},\ \bibinfo {pages} {7359} (\bibinfo {year}
  {1997})}\BibitemShut {NoStop}%
\bibitem [{\citenamefont {Held}(2007)}]{held2007electronic}%
  \BibitemOpen
  \bibfield  {author} {\bibinfo {author} {\bibfnamefont {K.}~\bibnamefont
  {Held}},\ }\href {https://doi.org/10.1080/00018730701619647} {\bibfield
  {journal} {\bibinfo  {journal} {Advances in Physics}\ }\textbf {\bibinfo
  {volume} {56}},\ \bibinfo {pages} {829} (\bibinfo {year} {2007})}\BibitemShut
  {NoStop}%
\bibitem [{\citenamefont {Rozenberg}\ \emph {et~al.}(1995)\citenamefont
  {Rozenberg}, \citenamefont {Kotliar}, \citenamefont {Kajueter}, \citenamefont
  {Thomas}, \citenamefont {Rapkine}, \citenamefont {Honig},\ and\ \citenamefont
  {Metcalf}}]{PhysRevLett.75.105}%
  \BibitemOpen
  \bibfield  {author} {\bibinfo {author} {\bibfnamefont {M.~J.}\ \bibnamefont
  {Rozenberg}}, \bibinfo {author} {\bibfnamefont {G.}~\bibnamefont {Kotliar}},
  \bibinfo {author} {\bibfnamefont {H.}~\bibnamefont {Kajueter}}, \bibinfo
  {author} {\bibfnamefont {G.~A.}\ \bibnamefont {Thomas}}, \bibinfo {author}
  {\bibfnamefont {D.~H.}\ \bibnamefont {Rapkine}}, \bibinfo {author}
  {\bibfnamefont {J.~M.}\ \bibnamefont {Honig}}, \ and\ \bibinfo {author}
  {\bibfnamefont {P.}~\bibnamefont {Metcalf}},\ }\href {\doibase
  10.1103/PhysRevLett.75.105} {\bibfield  {journal} {\bibinfo  {journal} {Phys.
  Rev. Lett.}\ }\textbf {\bibinfo {volume} {75}},\ \bibinfo {pages} {105}
  (\bibinfo {year} {1995})}\BibitemShut {NoStop}%
\bibitem [{\citenamefont {Pavarini}\ \emph {et~al.}(2005)\citenamefont
  {Pavarini}, \citenamefont {Yamasaki}, \citenamefont {Nuss},\ and\
  \citenamefont {Andersen}}]{Pavarini}%
  \BibitemOpen
  \bibfield  {author} {\bibinfo {author} {\bibfnamefont {E.}~\bibnamefont
  {Pavarini}}, \bibinfo {author} {\bibfnamefont {A.}~\bibnamefont {Yamasaki}},
  \bibinfo {author} {\bibfnamefont {J.}~\bibnamefont {Nuss}}, \ and\ \bibinfo
  {author} {\bibfnamefont {O.~K.}\ \bibnamefont {Andersen}},\ }\href
  {http://stacks.iop.org/1367-2630/7/i=1/a=188} {\bibfield  {journal} {\bibinfo
   {journal} {New Journal of Physics}\ }\textbf {\bibinfo {volume} {7}},\
  \bibinfo {pages} {188} (\bibinfo {year} {2005})}\BibitemShut {NoStop}%
\bibitem [{\citenamefont {Lichtenstein}\ \emph {et~al.}(2001)\citenamefont
  {Lichtenstein}, \citenamefont {Katsnelson},\ and\ \citenamefont
  {Kotliar}}]{PhysRevLett.87.067205}%
  \BibitemOpen
  \bibfield  {author} {\bibinfo {author} {\bibfnamefont {A.~I.}\ \bibnamefont
  {Lichtenstein}}, \bibinfo {author} {\bibfnamefont {M.~I.}\ \bibnamefont
  {Katsnelson}}, \ and\ \bibinfo {author} {\bibfnamefont {G.}~\bibnamefont
  {Kotliar}},\ }\href {\doibase 10.1103/PhysRevLett.87.067205} {\bibfield
  {journal} {\bibinfo  {journal} {Phys. Rev. Lett.}\ }\textbf {\bibinfo
  {volume} {87}},\ \bibinfo {pages} {067205} (\bibinfo {year}
  {2001})}\BibitemShut {NoStop}%
\bibitem [{\citenamefont {Pavarini}\ \emph {et~al.}(2004)\citenamefont
  {Pavarini}, \citenamefont {Biermann}, \citenamefont {Poteryaev},
  \citenamefont {Lichtenstein}, \citenamefont {Georges},\ and\ \citenamefont
  {Andersen}}]{pavariniPRL}%
  \BibitemOpen
  \bibfield  {author} {\bibinfo {author} {\bibfnamefont {E.}~\bibnamefont
  {Pavarini}}, \bibinfo {author} {\bibfnamefont {S.}~\bibnamefont {Biermann}},
  \bibinfo {author} {\bibfnamefont {A.}~\bibnamefont {Poteryaev}}, \bibinfo
  {author} {\bibfnamefont {A.~I.}\ \bibnamefont {Lichtenstein}}, \bibinfo
  {author} {\bibfnamefont {A.}~\bibnamefont {Georges}}, \ and\ \bibinfo
  {author} {\bibfnamefont {O.~K.}\ \bibnamefont {Andersen}},\ }\href {\doibase
  10.1103/PhysRevLett.92.176403} {\bibfield  {journal} {\bibinfo  {journal}
  {Phys. Rev. Lett.}\ }\textbf {\bibinfo {volume} {92}},\ \bibinfo {pages}
  {176403} (\bibinfo {year} {2004})}\BibitemShut {NoStop}%
\bibitem [{\citenamefont {Glazer}(1972)}]{Glazer:a09401}%
  \BibitemOpen
  \bibfield  {author} {\bibinfo {author} {\bibfnamefont {A.~M.}\ \bibnamefont
  {Glazer}},\ }\href {\doibase 10.1107/S0567740872007976} {\bibfield  {journal}
  {\bibinfo  {journal} {Acta Crystallographica Section B}\ }\textbf {\bibinfo
  {volume} {28}},\ \bibinfo {pages} {3384} (\bibinfo {year}
  {1972})}\BibitemShut {NoStop}%
\bibitem [{\citenamefont {Okimoto}\ \emph {et~al.}(1995)\citenamefont
  {Okimoto}, \citenamefont {Katsufuji}, \citenamefont {Okada}, \citenamefont
  {Arima},\ and\ \citenamefont {Tokura}}]{PhysRevB.51.9581}%
  \BibitemOpen
  \bibfield  {author} {\bibinfo {author} {\bibfnamefont {Y.}~\bibnamefont
  {Okimoto}}, \bibinfo {author} {\bibfnamefont {T.}~\bibnamefont {Katsufuji}},
  \bibinfo {author} {\bibfnamefont {Y.}~\bibnamefont {Okada}}, \bibinfo
  {author} {\bibfnamefont {T.}~\bibnamefont {Arima}}, \ and\ \bibinfo {author}
  {\bibfnamefont {Y.}~\bibnamefont {Tokura}},\ }\href {\doibase
  10.1103/PhysRevB.51.9581} {\bibfield  {journal} {\bibinfo  {journal} {Phys.
  Rev. B}\ }\textbf {\bibinfo {volume} {51}},\ \bibinfo {pages} {9581}
  (\bibinfo {year} {1995})}\BibitemShut {NoStop}%
\bibitem [{\citenamefont {Kresse}\ and\ \citenamefont
  {Furthmüller}(1996)}]{Kresse/Furthmueller_CMS:1996}%
  \BibitemOpen
  \bibfield  {author} {\bibinfo {author} {\bibfnamefont {G.}~\bibnamefont
  {Kresse}}\ and\ \bibinfo {author} {\bibfnamefont {J.}~\bibnamefont
  {Furthmüller}},\ }\href {\doibase
  https://doi.org/10.1016/0927-0256(96)00008-0} {\bibfield  {journal} {\bibinfo
   {journal} {Computational Materials Science}\ }\textbf {\bibinfo {volume}
  {6}},\ \bibinfo {pages} {15 } (\bibinfo {year} {1996})}\BibitemShut {NoStop}%
\bibitem [{\citenamefont {Kresse}\ and\ \citenamefont
  {Joubert}(1999)}]{Kresse/Joubert:1999}%
  \BibitemOpen
  \bibfield  {author} {\bibinfo {author} {\bibfnamefont {G.}~\bibnamefont
  {Kresse}}\ and\ \bibinfo {author} {\bibfnamefont {D.}~\bibnamefont
  {Joubert}},\ }\href {\doibase 10.1103/PhysRevB.59.1758} {\bibfield  {journal}
  {\bibinfo  {journal} {Phys. Rev. B}\ }\textbf {\bibinfo {volume} {59}},\
  \bibinfo {pages} {1758} (\bibinfo {year} {1999})}\BibitemShut {NoStop}%
\bibitem [{\citenamefont {Perdew}\ \emph {et~al.}(1996)\citenamefont {Perdew},
  \citenamefont {Burke},\ and\ \citenamefont {Ernzerhof}}]{PBE}%
  \BibitemOpen
  \bibfield  {author} {\bibinfo {author} {\bibfnamefont {J.~P.}\ \bibnamefont
  {Perdew}}, \bibinfo {author} {\bibfnamefont {K.}~\bibnamefont {Burke}}, \
  and\ \bibinfo {author} {\bibfnamefont {M.}~\bibnamefont {Ernzerhof}},\ }\href
  {\doibase 10.1103/PhysRevLett.77.3865} {\bibfield  {journal} {\bibinfo
  {journal} {Phys. Rev. Lett.}\ }\textbf {\bibinfo {volume} {77}},\ \bibinfo
  {pages} {3865} (\bibinfo {year} {1996})}\BibitemShut {NoStop}%
\bibitem [{\citenamefont {Dudarev}\ \emph {et~al.}(1998)\citenamefont
  {Dudarev}, \citenamefont {Botton}, \citenamefont {Savrasov}, \citenamefont
  {Humphreys},\ and\ \citenamefont {Sutton}}]{PhysRevB.57.1505}%
  \BibitemOpen
  \bibfield  {author} {\bibinfo {author} {\bibfnamefont {S.~L.}\ \bibnamefont
  {Dudarev}}, \bibinfo {author} {\bibfnamefont {G.~A.}\ \bibnamefont {Botton}},
  \bibinfo {author} {\bibfnamefont {S.~Y.}\ \bibnamefont {Savrasov}}, \bibinfo
  {author} {\bibfnamefont {C.~J.}\ \bibnamefont {Humphreys}}, \ and\ \bibinfo
  {author} {\bibfnamefont {A.~P.}\ \bibnamefont {Sutton}},\ }\href {\doibase
  10.1103/PhysRevB.57.1505} {\bibfield  {journal} {\bibinfo  {journal} {Phys.
  Rev. B}\ }\textbf {\bibinfo {volume} {57}},\ \bibinfo {pages} {1505}
  (\bibinfo {year} {1998})}\BibitemShut {NoStop}%
\bibitem [{\citenamefont {Lechermann}\ \emph {et~al.}(2006)\citenamefont
  {Lechermann}, \citenamefont {Georges}, \citenamefont {Poteryaev},
  \citenamefont {Biermann}, \citenamefont {Posternak}, \citenamefont
  {Yamasaki},\ and\ \citenamefont {Andersen}}]{lechermann2006dynamical}%
  \BibitemOpen
  \bibfield  {author} {\bibinfo {author} {\bibfnamefont {F.}~\bibnamefont
  {Lechermann}}, \bibinfo {author} {\bibfnamefont {A.}~\bibnamefont {Georges}},
  \bibinfo {author} {\bibfnamefont {A.}~\bibnamefont {Poteryaev}}, \bibinfo
  {author} {\bibfnamefont {S.}~\bibnamefont {Biermann}}, \bibinfo {author}
  {\bibfnamefont {M.}~\bibnamefont {Posternak}}, \bibinfo {author}
  {\bibfnamefont {A.}~\bibnamefont {Yamasaki}}, \ and\ \bibinfo {author}
  {\bibfnamefont {O.~K.}\ \bibnamefont {Andersen}},\ }\href {\doibase
  10.1103/PhysRevB.74.125120} {\bibfield  {journal} {\bibinfo  {journal} {Phys.
  Rev. B}\ }\textbf {\bibinfo {volume} {74}},\ \bibinfo {pages} {125120}
  (\bibinfo {year} {2006})}\BibitemShut {NoStop}%
\bibitem [{\citenamefont {Mostofi}\ \emph {et~al.}(2014)\citenamefont
  {Mostofi}, \citenamefont {Yates}, \citenamefont {Pizzi}, \citenamefont {Lee},
  \citenamefont {Souza}, \citenamefont {Vanderbilt},\ and\ \citenamefont
  {Marzari}}]{mostofi2014updated}%
  \BibitemOpen
  \bibfield  {author} {\bibinfo {author} {\bibfnamefont {A.~A.}\ \bibnamefont
  {Mostofi}}, \bibinfo {author} {\bibfnamefont {J.~R.}\ \bibnamefont {Yates}},
  \bibinfo {author} {\bibfnamefont {G.}~\bibnamefont {Pizzi}}, \bibinfo
  {author} {\bibfnamefont {Y.-S.}\ \bibnamefont {Lee}}, \bibinfo {author}
  {\bibfnamefont {I.}~\bibnamefont {Souza}}, \bibinfo {author} {\bibfnamefont
  {D.}~\bibnamefont {Vanderbilt}}, \ and\ \bibinfo {author} {\bibfnamefont
  {N.}~\bibnamefont {Marzari}},\ }\href {\doibase
  https://doi.org/10.1016/j.cpc.2014.05.003} {\bibfield  {journal} {\bibinfo
  {journal} {Comput. Phys. Commun.}\ }\textbf {\bibinfo {volume} {185}},\
  \bibinfo {pages} {2309 } (\bibinfo {year} {2014})}\BibitemShut {NoStop}%
\bibitem [{\citenamefont {Parcollet}\ \emph {et~al.}(2015)\citenamefont
  {Parcollet}, \citenamefont {Ferrero}, \citenamefont {Ayral}, \citenamefont
  {Hafermann}, \citenamefont {Krivenko}, \citenamefont {Messio},\ and\
  \citenamefont {Seth}}]{PARCOLLET2015398}%
  \BibitemOpen
  \bibfield  {author} {\bibinfo {author} {\bibfnamefont {O.}~\bibnamefont
  {Parcollet}}, \bibinfo {author} {\bibfnamefont {M.}~\bibnamefont {Ferrero}},
  \bibinfo {author} {\bibfnamefont {T.}~\bibnamefont {Ayral}}, \bibinfo
  {author} {\bibfnamefont {H.}~\bibnamefont {Hafermann}}, \bibinfo {author}
  {\bibfnamefont {I.}~\bibnamefont {Krivenko}}, \bibinfo {author}
  {\bibfnamefont {L.}~\bibnamefont {Messio}}, \ and\ \bibinfo {author}
  {\bibfnamefont {P.}~\bibnamefont {Seth}},\ }\href {\doibase
  https://doi.org/10.1016/j.cpc.2015.04.023} {\bibfield  {journal} {\bibinfo
  {journal} {Comput. Phys. Commun.}\ }\textbf {\bibinfo {volume} {196}},\
  \bibinfo {pages} {398 } (\bibinfo {year} {2015})}\BibitemShut {NoStop}%
\bibitem [{\citenamefont {Aichhorn}\ \emph {et~al.}(2016)\citenamefont
  {Aichhorn}, \citenamefont {Pourovskii}, \citenamefont {Seth}, \citenamefont
  {Vildosola}, \citenamefont {Zingl}, \citenamefont {Peil}, \citenamefont
  {Deng}, \citenamefont {Mravlje}, \citenamefont {Kraberger}, \citenamefont
  {Martins}, \citenamefont {Ferrero},\ and\ \citenamefont
  {Parcollet}}]{TRIQS/DFTTools}%
  \BibitemOpen
  \bibfield  {author} {\bibinfo {author} {\bibfnamefont {M.}~\bibnamefont
  {Aichhorn}}, \bibinfo {author} {\bibfnamefont {L.}~\bibnamefont
  {Pourovskii}}, \bibinfo {author} {\bibfnamefont {P.}~\bibnamefont {Seth}},
  \bibinfo {author} {\bibfnamefont {V.}~\bibnamefont {Vildosola}}, \bibinfo
  {author} {\bibfnamefont {M.}~\bibnamefont {Zingl}}, \bibinfo {author}
  {\bibfnamefont {O.~E.}\ \bibnamefont {Peil}}, \bibinfo {author}
  {\bibfnamefont {X.}~\bibnamefont {Deng}}, \bibinfo {author} {\bibfnamefont
  {J.}~\bibnamefont {Mravlje}}, \bibinfo {author} {\bibfnamefont {G.~J.}\
  \bibnamefont {Kraberger}}, \bibinfo {author} {\bibfnamefont {C.}~\bibnamefont
  {Martins}}, \bibinfo {author} {\bibfnamefont {M.}~\bibnamefont {Ferrero}}, \
  and\ \bibinfo {author} {\bibfnamefont {O.}~\bibnamefont {Parcollet}},\ }\href
  {\doibase http://dx.doi.org/10.1016/j.cpc.2016.03.014} {\bibfield  {journal}
  {\bibinfo  {journal} {Comput. Phys. Commun.}\ }\textbf {\bibinfo {volume}
  {204}},\ \bibinfo {pages} {200 } (\bibinfo {year} {2016})}\BibitemShut
  {NoStop}%
\bibitem [{\citenamefont {Seth}\ \emph {et~al.}(2016)\citenamefont {Seth},
  \citenamefont {Krivenko}, \citenamefont {Ferrero},\ and\ \citenamefont
  {Parcollet}}]{Seth2016274}%
  \BibitemOpen
  \bibfield  {author} {\bibinfo {author} {\bibfnamefont {P.}~\bibnamefont
  {Seth}}, \bibinfo {author} {\bibfnamefont {I.}~\bibnamefont {Krivenko}},
  \bibinfo {author} {\bibfnamefont {M.}~\bibnamefont {Ferrero}}, \ and\
  \bibinfo {author} {\bibfnamefont {O.}~\bibnamefont {Parcollet}},\ }\href
  {\doibase http://dx.doi.org/10.1016/j.cpc.2015.10.023} {\bibfield  {journal}
  {\bibinfo  {journal} {Comput. Phys. Commun.}\ }\textbf {\bibinfo {volume}
  {200}},\ \bibinfo {pages} {274 } (\bibinfo {year} {2016})}\BibitemShut
  {NoStop}%
\bibitem [{tag()}]{nota_dc}%
  \BibitemOpen
  \bibinfo {note} {{K}. Held, Advances in
  Physics {\bf 56}, 829 (2007). The formula for the double-counting correction
  is $E^{\rm dc}_i = \bar{U_i}(n_i-1/2)$, where $\bar{U}_i$ is the average
  Coulomb interaction felt by the orbital $i$: $\bar{U}=U-2J$ for the \tgg
  orbitals and $\bar{U}=U$ for the \ovv orbital.}\BibitemShut {Stop}%
\bibitem [{\citenamefont {Bryan}(1990)}]{Bryan1990}%
  \BibitemOpen
  \bibfield  {author} {\bibinfo {author} {\bibfnamefont {R.~K.}\ \bibnamefont
  {Bryan}},\ }\href {\doibase 10.1007/BF02427376} {\bibfield  {journal}
  {\bibinfo  {journal} {European Biophysics Journal}\ }\textbf {\bibinfo
  {volume} {18}},\ \bibinfo {pages} {165} (\bibinfo {year} {1990})}\BibitemShut
  {NoStop}%
\bibitem [{\citenamefont {Dymkowski}\ and\ \citenamefont
  {Ederer}(2014)}]{PhysRevB.89.161109}%
  \BibitemOpen
  \bibfield  {author} {\bibinfo {author} {\bibfnamefont {K.}~\bibnamefont
  {Dymkowski}}\ and\ \bibinfo {author} {\bibfnamefont {C.}~\bibnamefont
  {Ederer}},\ }\href {\doibase 10.1103/PhysRevB.89.161109} {\bibfield
  {journal} {\bibinfo  {journal} {Phys. Rev. B}\ }\textbf {\bibinfo {volume}
  {89}},\ \bibinfo {pages} {161109} (\bibinfo {year} {2014})}\BibitemShut
  {NoStop}%
\bibitem [{\citenamefont {Cwik}\ \emph {et~al.}(2003)\citenamefont {Cwik},
  \citenamefont {Lorenz}, \citenamefont {Baier}, \citenamefont {M\"uller},
  \citenamefont {Andr\'e}, \citenamefont {Bour\'ee}, \citenamefont
  {Lichtenberg}, \citenamefont {Freimuth}, \citenamefont {Schmitz},
  \citenamefont {M\"uller-Hartmann},\ and\ \citenamefont
  {Braden}}]{PhysRevB.68.060401}%
  \BibitemOpen
  \bibfield  {author} {\bibinfo {author} {\bibfnamefont {M.}~\bibnamefont
  {Cwik}}, \bibinfo {author} {\bibfnamefont {T.}~\bibnamefont {Lorenz}},
  \bibinfo {author} {\bibfnamefont {J.}~\bibnamefont {Baier}}, \bibinfo
  {author} {\bibfnamefont {R.}~\bibnamefont {M\"uller}}, \bibinfo {author}
  {\bibfnamefont {G.}~\bibnamefont {Andr\'e}}, \bibinfo {author} {\bibfnamefont
  {F.}~\bibnamefont {Bour\'ee}}, \bibinfo {author} {\bibfnamefont
  {F.}~\bibnamefont {Lichtenberg}}, \bibinfo {author} {\bibfnamefont
  {A.}~\bibnamefont {Freimuth}}, \bibinfo {author} {\bibfnamefont
  {R.}~\bibnamefont {Schmitz}}, \bibinfo {author} {\bibfnamefont
  {E.}~\bibnamefont {M\"uller-Hartmann}}, \ and\ \bibinfo {author}
  {\bibfnamefont {M.}~\bibnamefont {Braden}},\ }\href {\doibase
  10.1103/PhysRevB.68.060401} {\bibfield  {journal} {\bibinfo  {journal} {Phys.
  Rev. B}\ }\textbf {\bibinfo {volume} {68}},\ \bibinfo {pages} {060401}
  (\bibinfo {year} {2003})}\BibitemShut {NoStop}%
\bibitem [{\citenamefont {Eitel}\ and\ \citenamefont
  {Greedan}(1986)}]{EITEL198695}%
  \BibitemOpen
  \bibfield  {author} {\bibinfo {author} {\bibfnamefont {M.}~\bibnamefont
  {Eitel}}\ and\ \bibinfo {author} {\bibfnamefont {J.}~\bibnamefont
  {Greedan}},\ }\href {\doibase https://doi.org/10.1016/0022-5088(86)90220-1}
  {\bibfield  {journal} {\bibinfo  {journal} {Journal of the Less Common
  Metals}\ }\textbf {\bibinfo {volume} {116}},\ \bibinfo {pages} {95 }
  (\bibinfo {year} {1986})}\BibitemShut {NoStop}%
\bibitem [{tag()}]{nota_occupations}%
  \BibitemOpen
  \bibinfo {note} {{T}he DMFT occupations
  reported here correspond to the diagonal elements of the local Green's
  Function $-G(\tau)$ at $\tau = \beta$ for the corresponding site, which are
  defined in the so-called crystal-field basis where the local part of the
  non-interacting Hamiltonian is diagonal.}\BibitemShut {Stop}%
\bibitem [{\citenamefont {Karolak}\ \emph {et~al.}(2010)\citenamefont
  {Karolak}, \citenamefont {Ulm}, \citenamefont {Wehling}, \citenamefont
  {Mazurenko}, \citenamefont {Poteryaev},\ and\ \citenamefont
  {Lichtenstein}}]{Karolak2010}%
  \BibitemOpen
  \bibfield  {author} {\bibinfo {author} {\bibfnamefont {M.}~\bibnamefont
  {Karolak}}, \bibinfo {author} {\bibfnamefont {G.}~\bibnamefont {Ulm}},
  \bibinfo {author} {\bibfnamefont {T.}~\bibnamefont {Wehling}}, \bibinfo
  {author} {\bibfnamefont {V.}~\bibnamefont {Mazurenko}}, \bibinfo {author}
  {\bibfnamefont {A.}~\bibnamefont {Poteryaev}}, \ and\ \bibinfo {author}
  {\bibfnamefont {A.}~\bibnamefont {Lichtenstein}},\ }\href {\doibase
  https://doi.org/10.1016/j.elspec.2010.05.021} {\bibfield  {journal} {\bibinfo
   {journal} {Journal of Electron Spectroscopy and Related Phenomena}\ }\textbf
  {\bibinfo {volume} {181}},\ \bibinfo {pages} {11 } (\bibinfo {year}
  {2010})}\BibitemShut {NoStop}%
\end{thebibliography}%

\end{document}